\documentclass[9pt,twocolumn,twoside]{optica}
\setboolean{shortarticle}{false}
\setboolean{minireview}{false}

\title{Pitch-rotational manipulation of single cells and particles using single-beam thermo-optical tweezers}

\author[1]{Sumeet Kumar}
\author[1]{M. Gunaseelan}
\author[1]{Rahul Vaippully}
\author[1]{Amrendra Kumar}
\author[2]{Mithun Ajith}
\author[1]{Gaurav Vaidya}
\author[2]{Soumya Dutta}
\author[1,*]{Basudev Roy}

\affil[1]{Department of Physics, Indian Institute of Technology Madras, Chennai, 600036, India}
\affil[1]{Department of Electrical Engineering, Indian Institute of Technology Madras, Chennai, 600036, India}

\affil[*]{Corresponding author: basudev@iitm.ac.in}

\begin{abstract}
3D pitch rotation of microparticles and cells assumes importance in a wide variety of applications in biology, physics, chemistry and medicine. Applications such as cell imaging and injection benefit from pitch-rotational manipulation. Generation of such motion in single beam optical tweezers has remained elusive due to complicacies of generating high enough ellipticity perpendicular to the direction of propagation. Further, trapping an extended object at two locations can only generate partial pitch motion by moving one of the foci in the axial direction. Here, we use hexagonal-shaped upconverting particles and single cells trapped close to a gold-coated glass cover slip in a sample chamber to generate complete 360 degree and continuous pitch motion even with a single optical tweezers beam. The tweezers beam passing through the gold surface is partially absorbed and generates a hot-spot to produce circulatory convective flows in the vicinity which rotates the objects. The rotation rate can be controlled by the intensity of the laser light and the thickness of the gold layer. Thus such a simple configuration can turn the particle in the pitch sense. The circulatory flows in this technique have a diameter of about 5 $\mu$m which is smaller than those reported using acousto-fluidic techniques.
\end{abstract}

\setboolean{displaycopyright}{true}

\begin{document}

\maketitle

\section{Introduction}

Any rigid body can have six degrees of freedom, namely the three translational ones and the three rotational ones. Micromanipulation of these degrees of freedom can be done in a non-contact form using optical tweezers \cite{grier}, magnetic tweezers using magnetic properties of the material \cite{magnetic}, electrokinetic traps using the charge of the particles \cite{electrokinetic} and more recently using acousto-fluidic means \cite{huang1,huang2}. In all these techniques, the translational degrees of freedom can be well addressed while rotational degrees are not so well explored. Precise rotational manipulation of particles, cells and multicellular organisms impacts various disciplines including single-cell analysis \cite{hong,pusharsky}, drug discovery \cite{fraser} and organism studies \cite{caceres,rohde,coskun}. Distinct rotational behaviour due to different cell morphologies can, for example,  be used as a potential diagnostic method \cite{mohanty,elbez}. Many of the high resolution imaging techniques have a better transverse resolution than longitudinal resolution too \cite{superresolution}, making rotational manipulation very useful. 

Given that in the nomenclature of the airlines, the rotational degrees of freedom can be named the pitch \cite{pitch1}, the yaw \cite{halina,pitch2} and the roll, only the yaw has been explored in optical tweezers while the dielectrophoretic traps \cite{dielectrophoresis} can spin the particle in many ways but require complicated setups. Magnetic tweezers can spin the particle in all directions but lacks specificity, and cannot address an individual particle, relying upon the magnetic field in the entire region of interest, not to mention that only magnetic particles respond to it.  Acousto-fluidic means can spin the particle in the pitch sense but lacks specificity as well. It is here that a possible tool to spin the particle in the pitch sense using optical tweezers gains advantage retaining the specificity and yet controlling the motion well. 

Optical tweezers has been used as torque wrench to spin particles controllably in the yaw sense \cite{laporta}. However attempts at generating the pitch degree of rotational motion using tweezers have only been partially successful. One such attempt used two tweezers beams to simultaneously trap a large particle and pushing the focus of one of the tweezers beams deeper\cite{lokesh}. This can controllably generate pitch motion which is however not complete 360 degree motion. Other strategies involve usage of elliptically polarized light which, unfortunately, can never generate continuous pitch motion because the conventional techniques have so far been unable to generate high enough ellipticity \cite{misawa} in the pitch sense. 

It is in this context that we show a technique that can not only confine a single particle but also spin it in the pitch sense using thermo-optical effects. An infra-red laser beam is tightly focused into a sample chamber using a high numerical aperture (NA) objective lens. The bottom surface of the sample chamber has been coated with a layer of gold which partially absorbs the infra-red light and generate a hot-spot \cite{flores}. This hot-spot subsequently generates convection currents in the vicinity of the trapping region which then spins the trapped particle in the pitch sense. This technique is independent of the type of particle but only relies on the presence of a gold layer on bottom surface of the sample chamber. This configuration of the gold surface on cover slip of the sample chamber has never been explored before with a single beam optical tweezers.

\section{Mechanism of the process}

\begin{figure}[htbp]
\centering
\includegraphics[width=\linewidth]{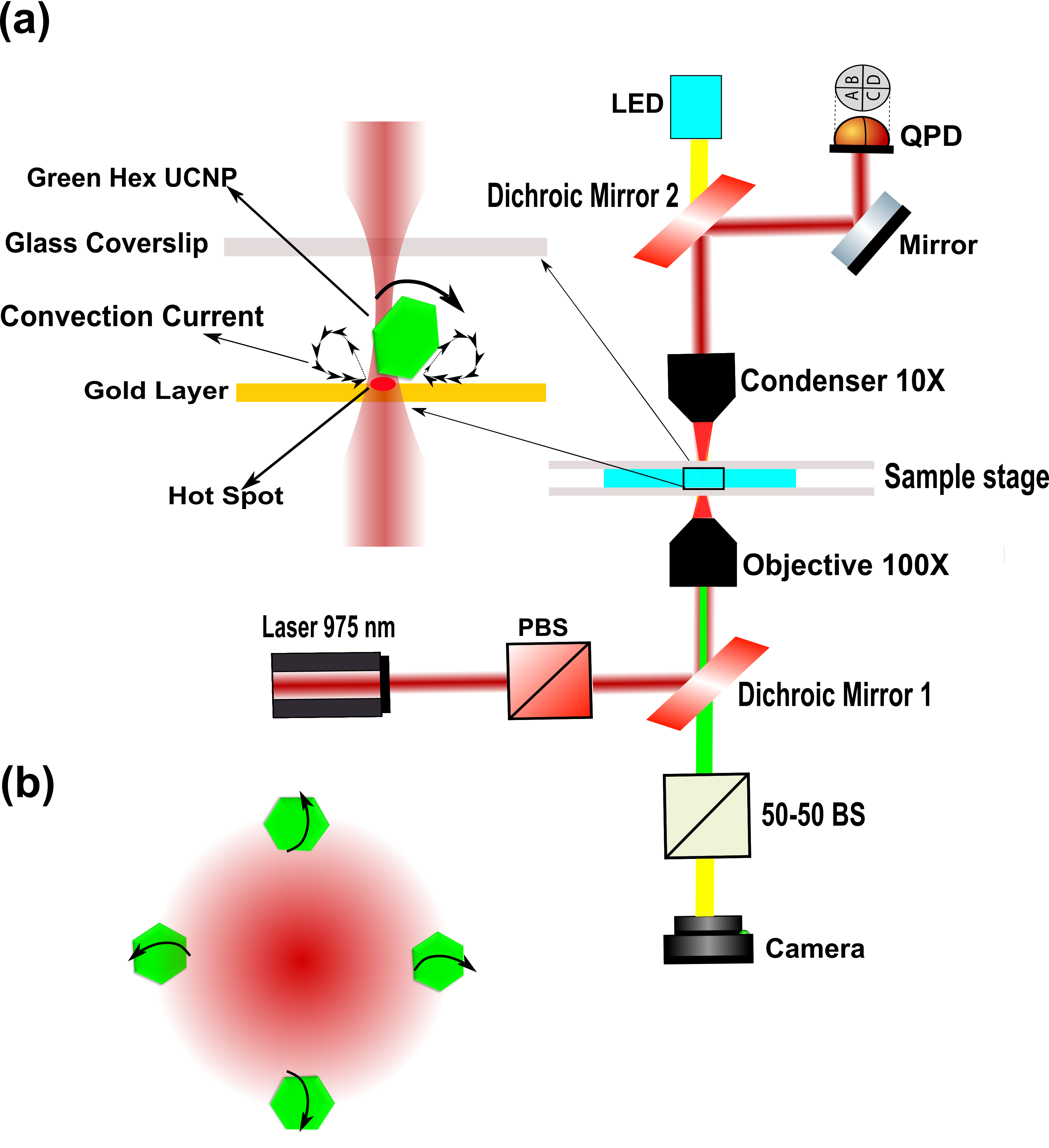}
\caption{(a) Schematic diagram of the apparatus. Shown in the inset is the formation of the hot-spot due to the presence of the trapping laser, and the subsequent pitch rotation of the particle (b) Direction of rotation of the particle at the sides of the hot-spot. }
\label{schematic}
\end{figure}

The Fig.\ref{schematic} shows a schematic diagram for the process. A sample chamber is made from a glass slide and a gold coated cover slip, with the gold surface at the bottom of the chamber. An laser beam is tightly focused by a high numerical aperture objective into the sample chamber such that the gold surface encounters the light first while in a converging geometry. Then the light focuses into the tweezers spot. This light is partially absorbed by the gold surface to create a hot spot. Once the hot-spot is formed, it induces circulatory convection currents in the water, such that water is directed radially inwards towards the spot close to the gold surface while rising up at the location of the spot and then going radially outwards at a plane away from the gold surface, as shown in Fig. \ref{schematic}(a). When we place either hexagonal shaped particles or even polystyrene particles suspended in water, inside the sample chamber, the tweezers light ensures that the particles come towards the hot spot where it gets captured by the convection currents and starts circulating in the pitch sense. 

The pitch motion is further helped by the fact that, hexagonal shaped up-converting nanoparticles (UCNP) are known to turn side-on when placed in a face-on configuration inside the tweezers beam\cite{rodriguez2016optical}. Once the beam turned side-on, the convection current close to the  gold surface pulls the edge towards the hot-spot while that at a higher plane pushes it radially away to generate a pitch motion and send the particle back into the face-on configuration. In this way, a stable complete and continuous pitch motion is generated.

We make a 2-dimensional simulation on COMSOL to understand the effect. We make a sample chamber configuration with a gold layer at the bottom surface. A region of 3 $\mu$m is illuminated with a
975 nm laser beam which is subsequently absorbed by the gold layer. The non iso-thermal flow (nitf) multi-physics module was used which includes
the laminar flow (spf) and heat transfer in fluids module (ht).
The fluid temperature was taken from the ht module and the velocity field
of the water and the pressure were taken from the nift module. The initial
temperature was set to 293.15K. Pair thermal contact was established between
the gold slab and the water sample. The gold slab was identified as a solid, to
distinguish its heat transfer properties from that of water. The laser source was
modeled using the deposited beam power boundary condition. A Gaussian beam
of varying power and 3µm spot size was made incident on the lower boundary
of the gold slab, as if it were originating from the coordinate $x=0$ m and $z= -0.2 m$ towards the positive z axis. All
surface to ambient losses were neglected for the purpose of this simulation. The
convection was modelled by considering convectively enhanced conductivity.
The thermal conductivity is increased by an empirical correlation factor that
depends on the dimensions of the water slab and the temperature variation
across the cavity. Even in doing so, the contributions of the velocity field of the
fluid were not ignored in the heat transfer module. A pressure point constraint was applied to the water. The pressure of the top
most layer was set to $10^5$ Pa. Gravity was considered in the negative z direction. The fluid was not
given an initial velocity field.

\begin{figure}[htbp]
\centering
\includegraphics[width=\linewidth]{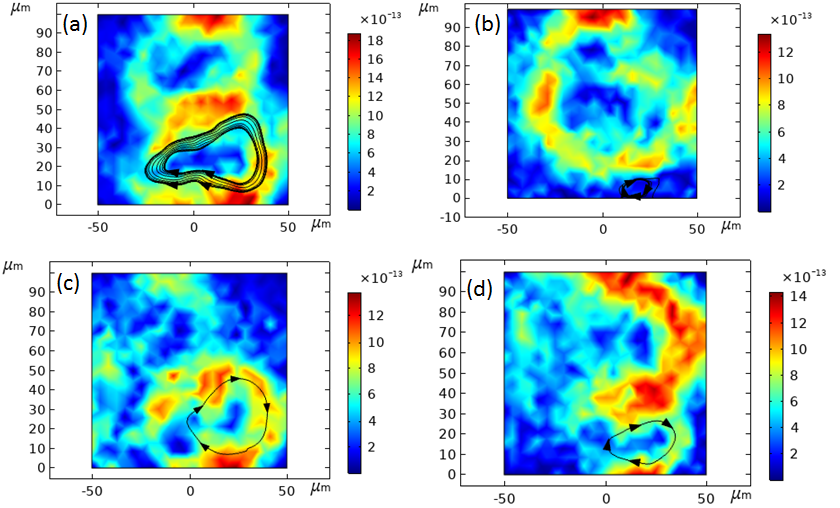}
\caption{Results of 2-D COMSOL simulations for a single laser beam incident via the bottom gold coated surface into the sample chamber at x=0 for a spot size of 3 $\mu$m. The laser propagates in z from 0 to 100 $\mu$m. (a) The thickness of the gold layer on glass sover slip is 15 nm, while 40 mW of laser light is made incident on the surface from below (b) Thickness of gold is 15 nm while 10 mW of laser light is incident (c) Thickness of gold is 20 nm while 40 mW of laser light is incident (d) Thickness of gold is 25 nm while 40 mW of laser light is incident. The color bar indicates the magnitude of the water velocity in m/sec. Arrows indicate the direction of water motion. }
\label{comsol}
\end{figure}

Then this generates the convection currents in water, as depicted in Fig. \ref{comsol}. The color code indicates the velocity of water at a certain region, indicated in m/sec while the direction of the fluid motion is indicated by the arrow. We find that there are circulatory regions appearing 20 $\mu$m away from the gold layer in the z direction. We also find that the circulatory region appear immediately above the hot spot for the thickness of the gold film to be 15 nm, as indicated in Fig. \ref{comsol}(a) at 40 mW of laser power, while it goes very close to the surface for a power of 10 mW, indicated in Fig. \ref{comsol}(b). The circulatory region moves to be side of the hot spot as the thickness of the gold layer is increased to about 20 nm (Fig. \ref{comsol}(c)) and 25 nm (Fig. \ref{comsol}(d)), with the incident power of the laser being 40 mW. We show circulatory regions on only one side of the hot-spot for convenience.

\section{Experimental details}

The experiment is performed using an inverted microscope configuration using an OTKB/M kit (Thorlabs, USA), as shown in Fig. \ref{schematic}. The bottom objective is a 100x, 1.3 NA Olympus oil-immersion objective while the condenser at the top is an E Plan 10X, 0.25 NA air-immersion one from Nikon. A diode laser at 1064 nm wavelength (Lasever, 1.5 W maximum power) or a butterfly laser at 975 nm wavelength (Thorlabs, USA, 300 mW maximum power) have been used for the experiment. The laser passes Dichroic Mirror 1 at the bottom to be directed into the sample chamber whereafter, the forward scattered light is collected using the condenser and directed via the Dichroic Mirror 2 into the Quadrant Photodiode (QPD, Thorlabs, USA). The data acquired by the QPD is sent into the computer via a data acquisition card (NI PCI 6143). A white light LED was used to illuminate the sample from the top through the Dichroic Mirror 2 which is eventually collected through the Dichroic Mirror 1 at the bottom and sent into a CMOS camera (Thorlabs, USA). 

The sample chamber is constructed from a gold coated cover slip (Blue Star, number 1 size, english glass) with three values of thicknesses, namely, 25 nm, 35 nm and 100 nm. The top surface is a glass slide (Blue Star, 75 mm length, 25 mm width and a thickness of 1.1 mm) and the sample containing particles dispersed in water placed in the middle. 

For coating the gold, the glass substrate was cleaned in acetone, IPA and de-ionized water using an ultrasonic bath for 5 min each and followed by nitrogen drying.The samples were loaded into a thermal evaporator immediately and 5 nm chromium was first evaporated at a rate of 0.05 nm/sec to provide better adhesion followed by 30 nm gold evaporation at 0.1 nm/sec. The thickness of deposition was monitored using a quartz crystal monitor and the vacuum of the evaporator chamber was maintained at 1 nbar throughout the deposition.

Typically, the particle used is that of 5 $\mu$m length, hexagonal shaped, upconverting nanoparticles of $NaYF_4:Yb,Er$, or a dispersion of 1 $\pm$ 0.05 $\mu$m diameter polystyrene spheres (Thermo Fisher Scientific) in water. The $NaYF_4:Yb,Er$ microcrystals were prepared by the conventional hydrothermal method \cite{Patricia} with certain modifications. A quantity of 1.2548 g of $Y(NO_3)_3$ and 1.2321 g of sodium citrate were added to 14 ml of water and vigorously stirred for about 10 min. Then, 21 ml of the aqueous solution of $Yb(NO_3)_3$ (0.3773 g) and $Er(NO_3)_3$ (0.0373 g) were mixed into the above mentioned solution. The milky white solution subsequently transformed itself into a transparent solution upon adding 67 ml of aqueous solution of NaF (1.411 g). The resultant solution was then transferred to a teflon lined stainless steel vessel (200 ml) and heated at 200 degree C for 12 h. The solution was washed with ethanol and water for four times upon cooling to room temperature. Finally, the white powder was collected by calcination at 100 degree C for 12 h. This powder was mixed with de-ionised water to make the upconverting nanoparticle suspension. 

For the prepration of the dictyostelium cells, wild-type D. Discoideum AX2 cells were obtained by inoculating the spores in 90 mm tissue culture plates containing axenic HL5 growth medium (HLG01XX - Formedium, Norfolk, UK, pH 6.4) supplemented with penicillin (100 units/mL) and streptomycin sulphate (100 mg/mL). The cultures were incubated at 22 degree C till semi-confluent plates were obtained. The mid-log cells were harvested in ice cold KK2 buffer (2.2 g/L $KH_2PO_4$ and 0.7 g/L $K_2HPO_4$, pH 6.4), washed twice and 5 $\times$ 106 cells were pelleted and placed in ice for performing further tests.

\section{Results and discussions}

\begin{figure*}[htbp]
\centering
\includegraphics[width=12 cm]{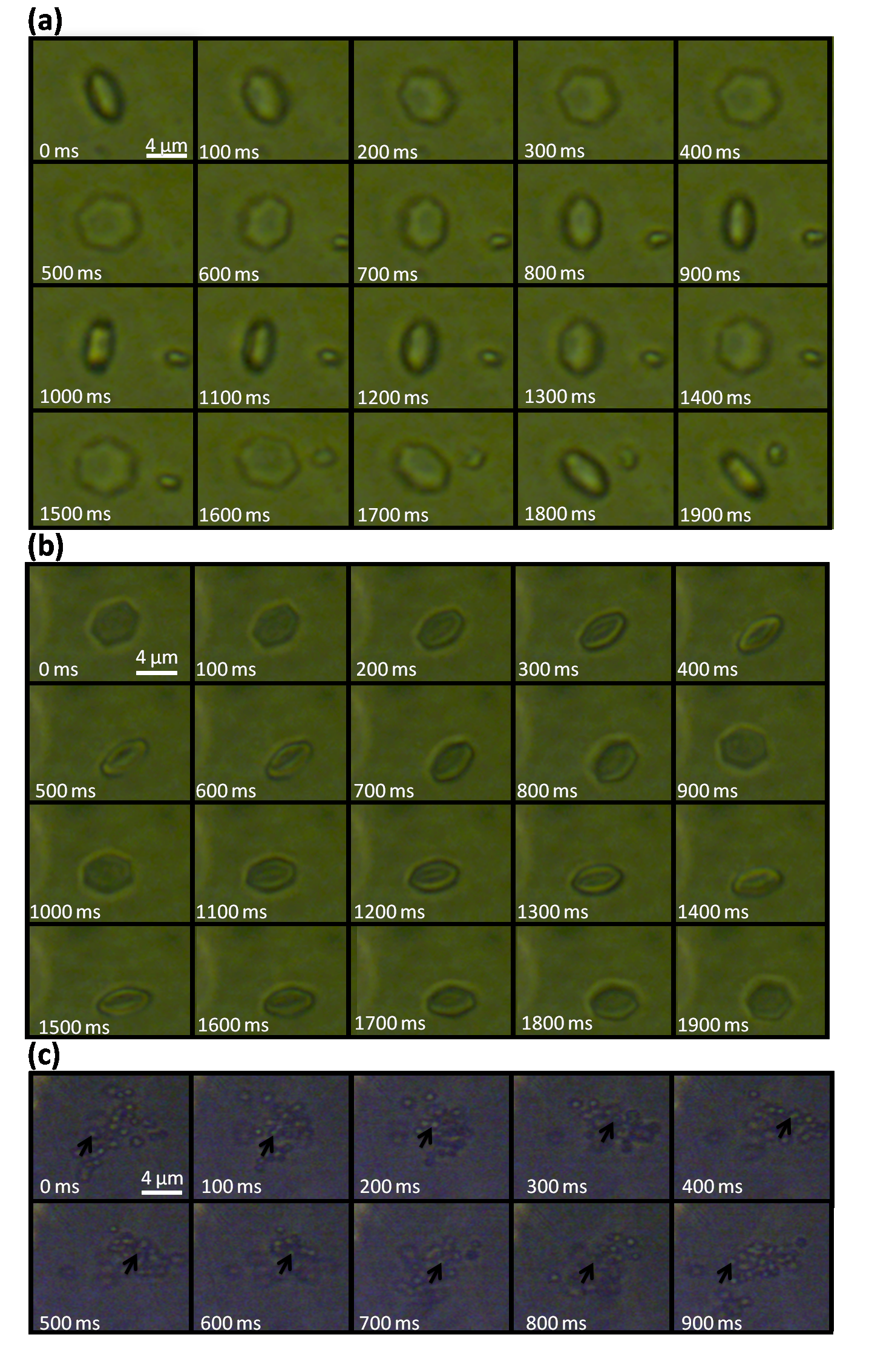}
\caption{Pitch rotation of perticles (a), (b) The hexagonal shaped UCNP spinning, (c) A cluster of polystyrene particles stably confined and spinning due to convective flows. The thickness of the gold layer was 30 nm while the laser power was about 10 mW. }
\label{pitch_motion}
\end{figure*}

\begin{figure}[htbp]
\centering
\includegraphics[width=\linewidth]{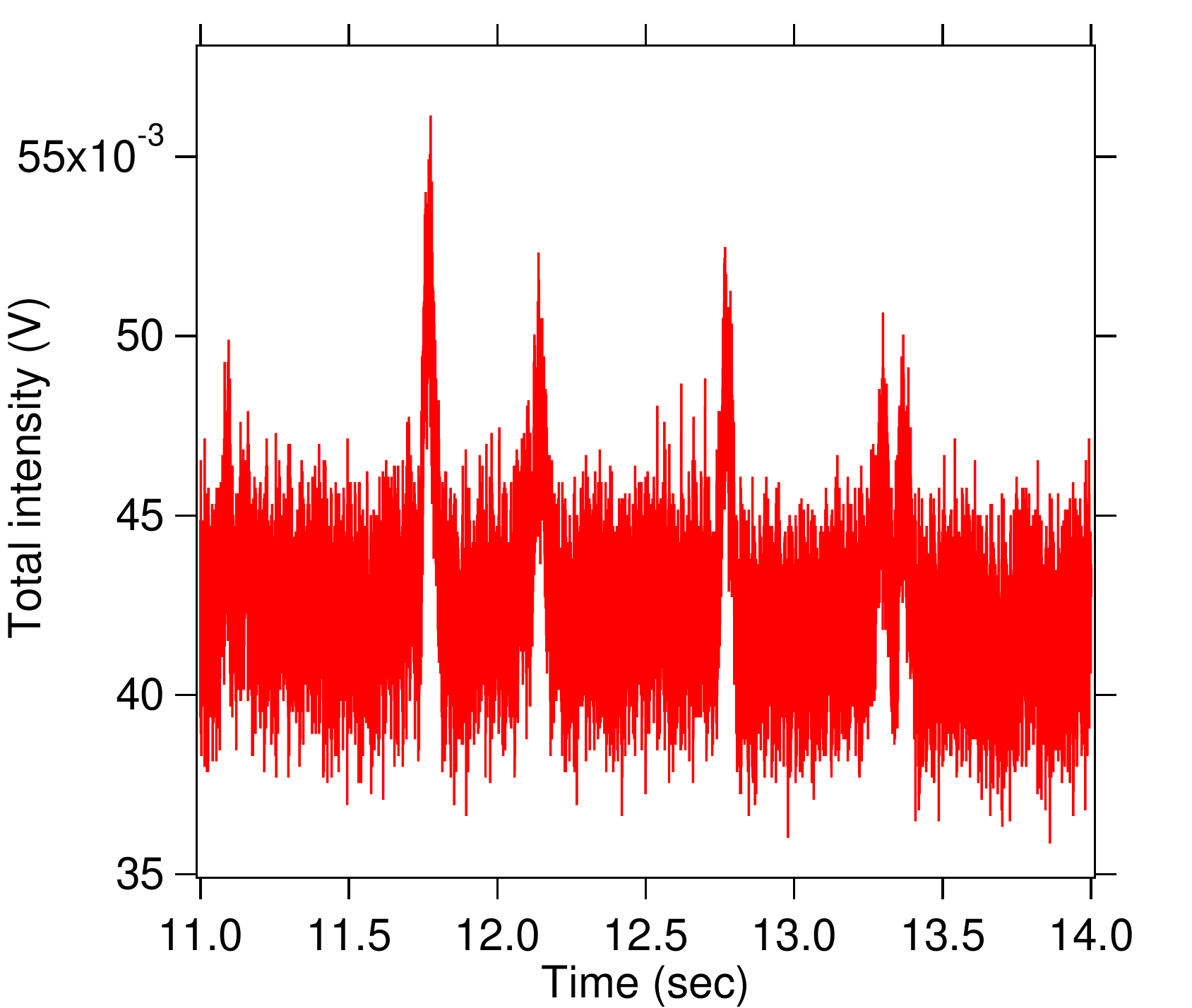}
\caption{ This figure indicates the time series for the total intensity of forward scattered light while the UCNP particle is executing pitch motion.  }
\label{time_series}
\end{figure}

We find that when the hexagonal UCNP particles are placed in such a configuration, the particles start to spin while being stably confined in three-dimensions, as shown in Fig. \ref{pitch_motion} (a) and (b). The particle is confined in a region immediately above the hot-spot in Fig. \ref{pitch_motion}(a), for a gold layer thickness of 25 nm, while it shifts to the side of the hot-spot for higher thickness of gold. The sense of rotation can be controlled by placing the particle slightly to the side of the hot-spot. Using the 975 nm laser for trapping the particle, the wavelength at which the particle has an absorption resonance\cite{GUNASEELAN2018174}, the emission is found to be completely absent. We believe this to be due to absorption of the emitted visible light by the gold film, which might accentuate the process of heating at the hot-spot. Nonetheless, the rotation is still observed for 1064 nm wavelength laser which is not at the absorption resonance and does not cause much emission. The circulatory flows in this technique have a diameter of about 5 $\mu$m which is smaller than those reported using acousto-fluidic \cite{huang1} techniques. Fig. \ref{time_series} indicates the total intensity of the forward scattered light as the particle executes pitch motion. The sudden changes in intensity are visible in this time series.

Similar circulatory effects are also observed when a suspension of polystyrene particles in water is used, as shown in Fig. \ref{pitch_motion} (c). Thus the effect is mainly due to the formation of the hot-spot by absorption of the laser beam by gold, referred to as the thermo-optic tweezers. This can confine the particles in three dimensions. 

Such circulations were believed to be formed deep inside the sample chamber, and never observed previously. The main difference in our set-up is the presence of the gold surface as opposed to any other dielectric material that causes these circulatory currents to be developed. 

\begin{figure}[htbp]
\centering
\includegraphics[width=\linewidth]{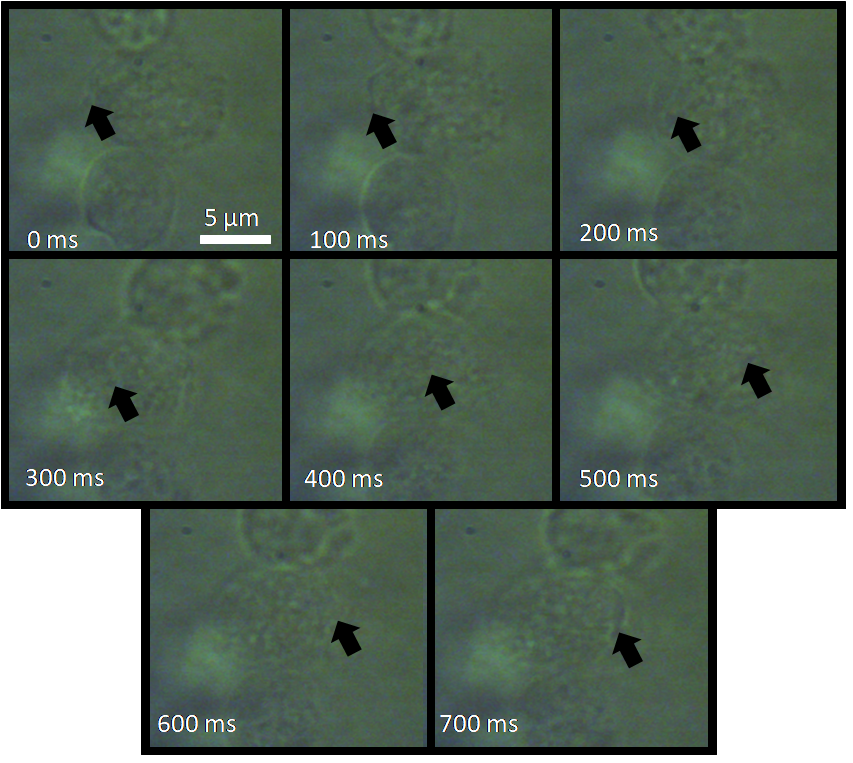}
\caption{The sequence of images shows the pitch rotation of an individual dictyostelium cell as performed with thermo-optical tweezers.}
\label{dicty}
\end{figure}

We use the same configuration to confine and rotate dictyostelium cells, as shown in Fig \ref{dicty}. The hot-spot in this case is immediately to the right of the cell. Thus we can turn individual live cells. 

The rate of rotation of the particle or the object is directly controlled by the rate of circulation of the water. We find that the rate of rotation is dependent upon the thickness of the gold layer. At low values of thickness, the rate of rotation is low, while it tends to increase when cover slips with thicker gold layers are used. The thicker gold layer absorbs more light leading to a higher temperature at the hot-spot and subsequently a higher velocity of the water circulating. Further, we also find a similar effect with laser power.

\begin{figure}[htbp]
\centering
\includegraphics[width=\linewidth]{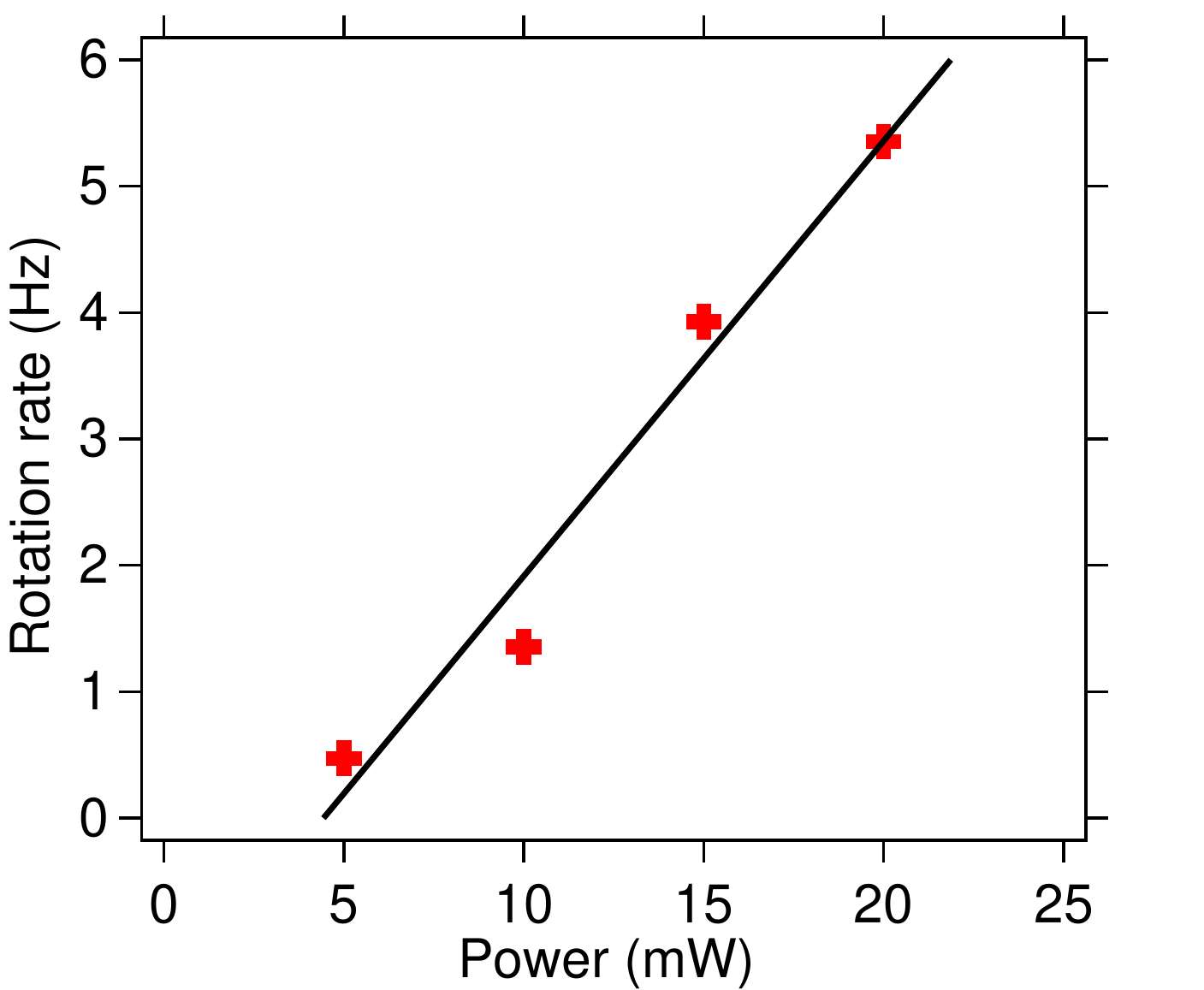}
\caption{Variation of rotation rate of the UCNP particle as a function of laser power.}
\label{rotation}
\end{figure}

Increasing the power generates higher temperature at the hot-spot and subsequently generates higher temperature gradients leading to higher velocity of water, shown in Fig. \ref{rotation}. 

\section{Conclusions}

Thus, to conclude, we have been able to generate circulatory convection currents in a sample chamber using thermo-optical tweezers where a gold surface is placed on the bottom surface. This allows the circulatory currents to be formed very close to the gold surface and hence is able to spin any trapped particle in the pitch sense. This is important, as generation of the pitch motion in single-beam optical tweezers has not been possible. Such pitch rotation has also been shown to turn individual dictyostelium cells which can then be used for imaging or injection purposes. Moreover, this rotation of the particle can be controlled by changing the amount of laser light incident in the trapping region, and the thickness of the gold film placed on the bottom cover slip. 

\section*{Acknowledgments}

We thank the Indian Institute of Technology, Madras for their seed and initiation grants. We also thank R. Baskar for the dictyostelium cells and J. Senthilselvan for the help with making upconverting nanoparticles.


\bibliography{sample}

\bibliographyfullrefs{sample}


\end{document}